\documentclass[%
 reprint,
 amsmath,amssymb,
 aps,
]{revtex4-2}

\usepackage{graphicx}
\usepackage[english]{babel}
\usepackage{dcolumn}
\usepackage{bm}
\begin{document}

\preprint{APS/123-QED}

\title{Divergence control of high-harmonic generation}

\author{Sylvianne Roscam Abbing$^{1,*}$, Filippo Campi$^1$, Faegheh S. Sajjadian$^1$, Nan Lin$^2$, Peter Smorenburg$^2$, Peter M. Kraus$^{1,*}$ \vspace{0.3cm}}

\affiliation{ $^1$Advanced Research Center for Nanolithography, Science Park 106, 1098 XG Amsterdam, The Netherlands\\ $^2$ASML Research, ASML Netherlands B.V., 5504 DR Veldhoven, The Netherlands \vspace{0.3cm}}

\email{Corresponding authors: roscam@arcnl.nl, kraus@arcnl.nl}

\date{\today}

\begin{abstract}
We show that the divergence of extreme ultraviolet pulses from high harmonic generation, which is directly linked to the shape and size of the refocused beam, can be controlled by the relative delay between the fundamental and its intense orthogonally polarized second harmonic in two-color high-harmonic generation. We find that the divergence is minimized close to the delays were the number of emitted photons is maximized. These findings are rationalized as suppression and enhancement of long and short electron trajectories as function of the phase of the two-color laser field, respectively. The orthogonally polarized second harmonic introduces a lateral momentum component that can select one trajectory whereas it deflects the other. At the same time, the second harmonic profoundly influences the tunnel ionization process that initiates high-harmonic generation, which provides another trajectory gate. Our scheme for controlling the divergence facilitates larger numerical apertures for extreme ultraviolet microscopy. In addition, the associated reduction of the size of the focus is beneficial for extreme ultraviolet nonlinear optics and spectroscopy, as well as imaging and metrology of embedded structures.
\end{abstract}


\maketitle


\section{Introduction}

\begin{figure*}[htbp]
\centering
\includegraphics[width=0.8\linewidth]{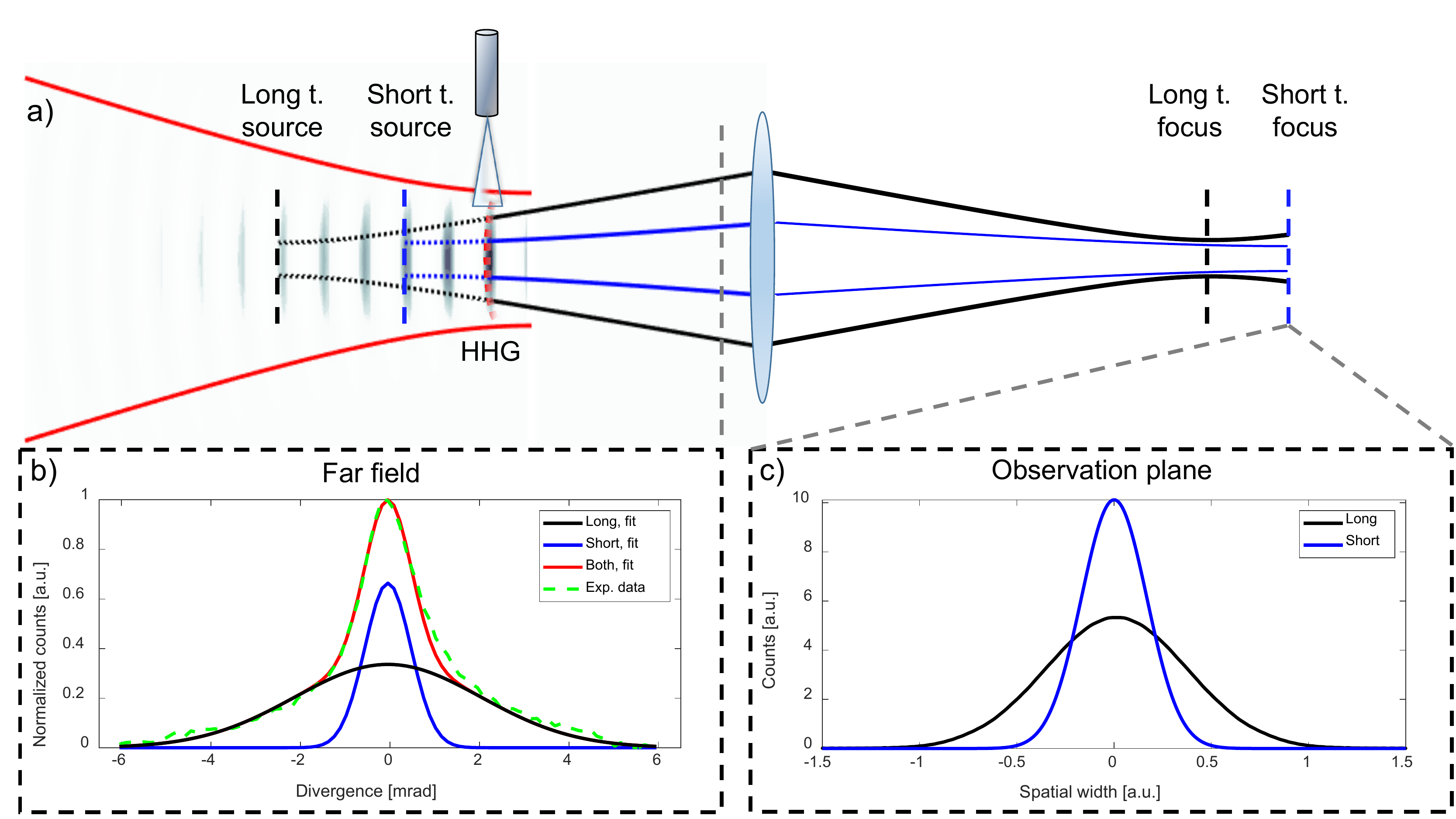}
\caption{\textbf{Virtual sources and foci in HHG:} (a) Schematic representation of an HHG source and refocusing of the harmonics (a lens is shown for simplicity). For a given harmonic order \textit{q}, the different contributions of the dipole phases for the short and the long trajectories give rise to different positions of the corresponding virtual sources (dashed blue and black lines). When the harmonics are refocused, the image of the long trajectories will be out of focus, leading to an overall smearing of the profile in the observation plane. (b) Experimental (green dashed line) spectrally-integrated, divergence-resolved intensity profile in the far field, fitted with a double Gaussian distribution (red solid line), representing the contributions of short (blue line) and long (black line) trajectories, respectively. (c) Intensity distribution in the observation (focal) plane, obtained by Fourier propagation. An additional quadratic radial phase distribution was added to the profile for the long trajectories, in order to estimate the different divergence and virtual source position.
}
\label{fig:prob}
\end{figure*}

High-harmonic generation (HHG) \cite{mcpherson87a,ferray88a} is the cornerstone of attosecond science \cite{corkum07a,krausz09a,kraus18a,kraus18b}, and equally important for table-top coherent diffraction applications \cite{sandberg07a,witte14a,zuerch14a,gardner17a}.
Many spectroscopy, imaging and possibly industrial metrology applications \cite{boef16a,kinoshita14a} strictly require small focus sizes with limited energy in the wings, which would otherwise produce diffraction artifacts from metrology targets that are oftentimes embedded into an integrated-circuit infrastructure \cite{boef16a}. Furthermore, larger divergent beams reduce the numerical aperture of an optical system.\\
Micro-focusing of broadband pulses from HHG can be achieved by ellipsoidal \cite{motoyama16a} or multiple toroidal grazing incidence mirrors \cite{poletto13a,coudert-alteirac17a}, and subsequent focus \cite{valentin03a} and wavefront sensing \cite{freisem18a} for optimizing optics alignment.
However, the quantum nature of the HHG process poses limitations to the focusing of the pulses from HHG: When an electron is ionized in a strong laser field, the electron accumulates an intensity-dependent phase during  propagation in the continuum, which is imprinted on the emitted high harmonics following electron-ion recombination \cite{krause92a,corkum93a,lewenstein94a}. The typical Gaussian intensity distribution (Fig.~\ref{fig:prob}(a)) of the driving field in focus therefore gives rise to a double-Gaussian divergence-resolved profile of HHG (Fig.~\ref{fig:prob}(b),  \cite{lewenstein95a,bellini98a,gaarde99a,schapper10a}), because each emitted harmonic can be ascribed to a set of two discrete quantum paths, the short and long electron trajectories. These trajectories have a unique transit time of the continuum electron and thus a well defined dipole phase that is imprinted on the emission.

This origin of divergence in HHG has been understood in a number of early seminal publications \cite{bellini98a,gaarde99a,salieres01a,lewenstein94a,lewenstein95a}.
However, it only became clear very recently that the different trajectories give rise to aberrations, which hinder refocusing the pulses from HHG to high intensities \cite{wikmark19a,quintard19a}. 
This is caused by the different virtual sources of long and short trajectories even if they correspond to the same photon energy (illustrated in Fig.~\ref{fig:prob}(a)). Therefore refocusing pulses from HHG does not give a clean Gaussian profile but rather a double Gaussian focus, where long and short trajectories effectively have different focal planes (Fig.~\ref{fig:prob}(c)). This makes a strong demagnification of the focused beam challenging.
The situation becomes worse when considering the broadband spectra associated with HHG, with every harmonic order and every trajectory having a different virtual source, which gives rise to strong chromatic aberrations of HHG pulses \cite{wikmark19a,quintard19a}. The double-Gaussian far-field divergence, just like the different virtual sources, are both phenomena that arise due to the existence of both long and short trajectories. Consequently, robust methods to control the divergence are needed, which should simultaneously maintain or improve the overall flux.\\
We demonstrate that two-color HHG can minimize the divergence and improve the flux at the same time by tuning the relative phase of the two-color field to selectively enhance the short over the long trajectory emission.
\\
Our work in this article builds on a number of earlier papers on two-color HHG. 
Two-color HHG introduces an additional momentum to the strong-field driven electron in the continuum. This approach enabled the reconstruction of electron transit times \cite{dudovich06a,shafir12a} using a weak second harmonic for both short and long trajectories \cite{soifer13a}. Two-color HHG with a strong second harmonic \cite{he10a} enabled selecting long and short electron trajectories \cite{brugnera11a}. However, it remained unclear if and how much this strategy would change the divergence, possibly at the expense of overall conversion efficiency. In a separate work, two-color HHG was shown to improve the flux of plateau harmonics dramatically \cite{kim05a}. However, in this study the cutoff was not detected by the spectrometer, which makes any quantitative assessment of the total flux difficult.

\section{Experimental concept and setup}
We use an 800 nm Titanium:Sapphire (Ti:Sa) laser with 40 fs pulse duration and 1 kHz repetition rate in our experiments.
The orthogonally polarized two-color field is generated with a 0.2 mm thick $\beta$-bariumborate (BBO) crystal that produces up to 25\% relative intensity of the second harmonic. Subsequently, two calcite plates compensate for the group delay between the 800~nm and 400~nm pulses, and a pair of fused silica wedges is used for fine adjustment of the group and phase delay of the two-color field. Both calcite plates were tilted slightly off-normal with opposite angles to compensate for possible spatial walkoffs between the fundamental and second harmonic due to refraction, which would lead to slightly nonlinear geometries and therefore smear out the beam profiles. One calcite plate was motorized to scan the relative two color phase delay. We had to scan for less than 0.3 degrees to achieve a 2$\pi$ two-color phase shift.\\
The two-color pulses were focused with f=50 cm focal length to intensities of $(1.8\pm0.3)\cdot10^{14}$~W/cm$^2$ into an effusive gas expansion of Ar. The nozzle was connected to an xyz-translation stage for fine positioning of the focus relatively to the gas. Subsequently, the HHG beam was spectrally dispersed with a concave abberation-corrected flat-field grating onto a double-stack microchannel plate detector backed with a phosphor screen. This geometry serves as far-field spectrometer, where the spectrum is dispersed in the horizontal plane, and the beam freely propagates in the vertical direction.
All results are systematically compared to HHG with 800~nm only, obtained by detuning the BBO to eliminate all second-harmonic conversion. In addition, we verified that the pulse energies of the 800 nm pulses compared to the 800+400~nm pulses were identical. For these settings, the cutoffs for HHG with 800 nm only and 800+400 nm were the same or 800~nm only produced one harmonic order more. We found that the exact intensity did not influence the divergence a lot, but the overall flux of the HHG. Therefore our estimates for the flux improvement with two-color pulses are on purpose very conservative, and larger factors might be achievable.\\

\begin{figure}[htbp]
\centering
\includegraphics[width=\linewidth]{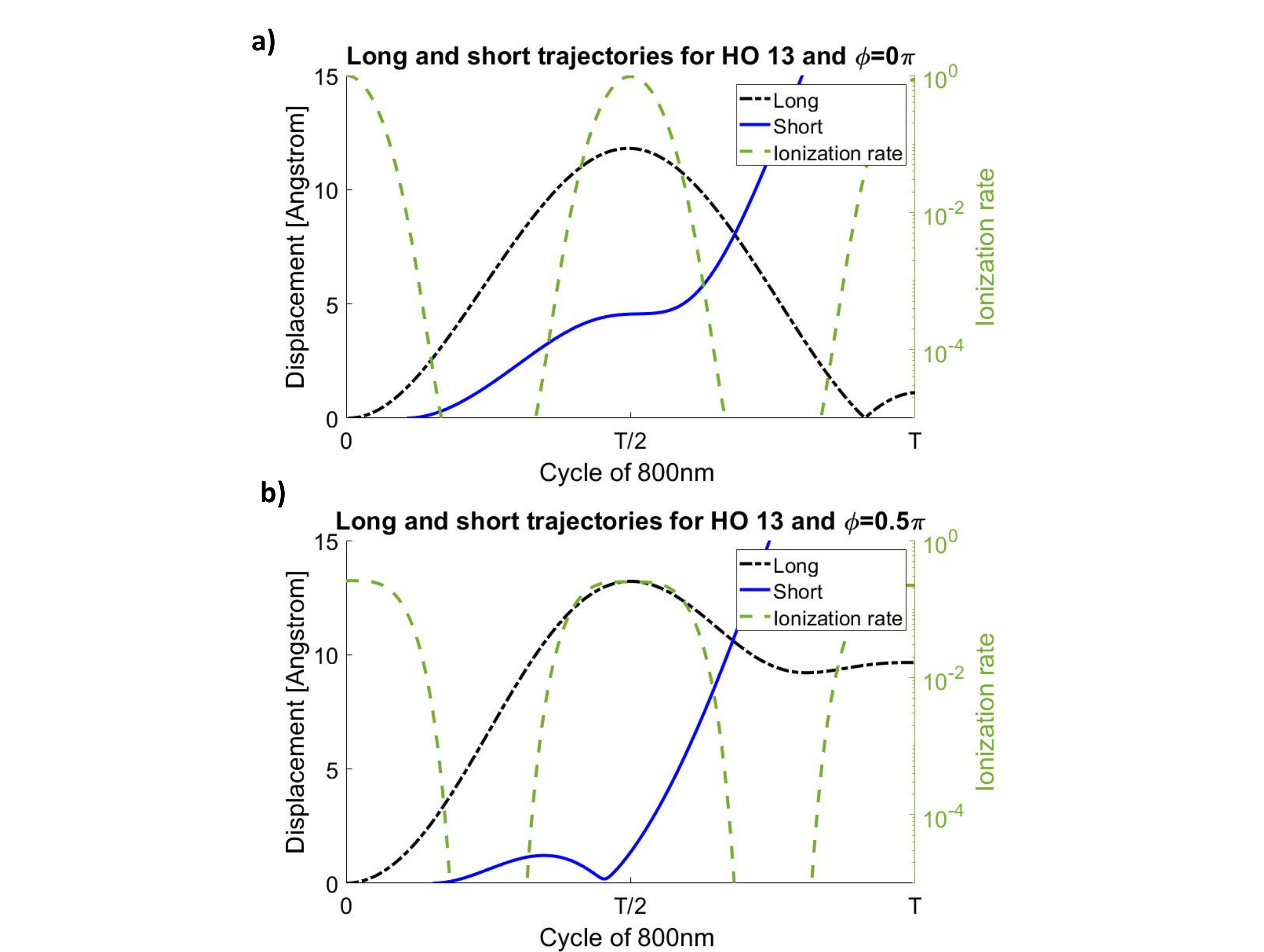}
\caption{\textbf{Concept of trajectory-control in two-color HHG:} The short (blue) and long (black) trajectories for harmonic 13 (61 nm) as well as the two-color ionization rate (green) are shown for a relative two-color phase of (a) 0 and (b) 0.5$\pi$~rad. For a phase of 0 rad in (a), the short trajectory misses the origin and the long trajectory recombines. The situation is vice versa in (b). In addition the relative difference in ionization rates between long and short trajectories is much less severe in (b) than in (a).}
\label{fig:solu}
\end{figure}

The concept of the experiment is illustrated in Fig.~\ref{fig:solu}, which shows the classically calculated long and short electron trajectories of H.O.~13 (61nm) in an orthogonally polarized two-color laser field for two different two-color phases in (a) and (b). The figures also show the relative instantaneous strong-field ionization rate in the two-color field \cite{keldysh65a,yudin01a}.
Many groups have previously studied the effect of an low-intensity cross polarized second harmonic on HHG: only when the field is chosen such that the electron is returning to its origin at the time of return, recombination is efficient. That occurs when the initial lateral velocity is compensated by the lateral displacement introduced by the second harmonic  \cite{shafir12a}. However, in the present article the field of the second harmonic is strong, so in addition the second harmonic has a profound influence on the strong-field-ionization step. 
For 0~rad phase (Fig.~\ref{fig:solu}(a)), the long trajectory recombines to the origin, and the short trajectory misses the origin. In addition, the ionization rate at the time of birth of the short trajectory is about 1000 times lower than at the maximum of the pulse. For a phase of 0.5$\pi$ ~rad, the short trajectory recombines, and the long trajectory misses its ion. Moreover, the ionization rate at the time of birth of the short trajectory is only 400 times lower than at the maximum of the pulse. So both ionization and recombination favor the short trajectory for this two-color phase. \\
While the steering of the continuum trajectories will influence the divergence through trajectory selection, the modulation in the ionization step will have a profound influence on the overall HHG flux. Orthogonally polarized two-color drivers were previously reported to cause a higher CE than in conventional one-color HHG \cite{kim05a}. We will show here that these settings of maximum CE correspond to a two-color phase that produces HHG with close-to minimum divergence, due to selection of short trajectories over long trajectories.

\section{Results and Discussion}
\begin{figure*}[htbp]
\centering
\includegraphics[width=0.8\linewidth]{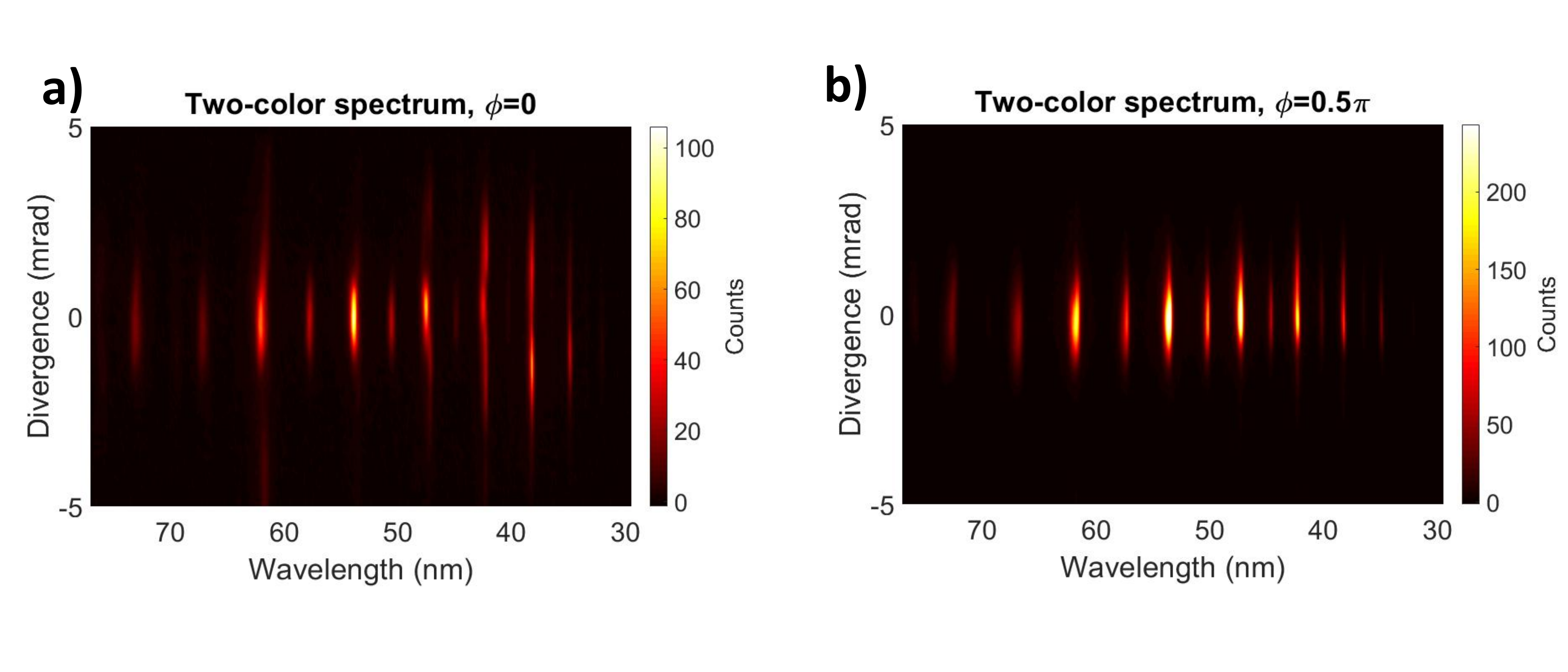}
\caption{ \textbf{Far-field HHG spectra for two relative two-color phases:} a) HHG spectrum for a relative phase of 0, in which long trajectories are selected. b) HHG spectrum for a relative phase of 0.5$\pi$, in which short trajectories are selected. The spectrum in (b) shows less signal at higher divergences, higher counts in the central area, and higher overall counts as compared to the spectrum in (a).}
\label{fig:spec}
\end{figure*}

Two-color HHG spectra for two different phases (0 and 0.5$\pi$) of the two-color field are shown in Fig.~\ref{fig:spec} as a function of both wavelength and divergence. A relative two-color phase delay of $\pi$ corresponds to a delay of about 670~as. In general, two-color HHG leads to the emission of both odd and even harmonics orders due to the broken inversion symmetry of the field. For the HHG conditions in Fig.~\ref{fig:spec}(a) we observe a strongly divergent spectrum, whereas the beam is more collimated for a two color-phase of 0.5$\pi$, as shown in Fig.~\ref{fig:spec}(b). In addition, the overall number of emitted photons is considerably higher for two-color phases where a more collimated beam is generated (Fig.~\ref{fig:spec}(b)). As described and simulated in Fig.~\ref{fig:solu}, a two-color phase of 0 corresponds to selecting predominantly long trajectories, which explains the more divergent profile in Fig.~\ref{fig:spec}(a), whereas a phase of 0.5$\pi$ selects short trajectories leading to a less divergent beam profile.
\\
\begin{figure*}
\centering
\includegraphics[width=0.9\linewidth]{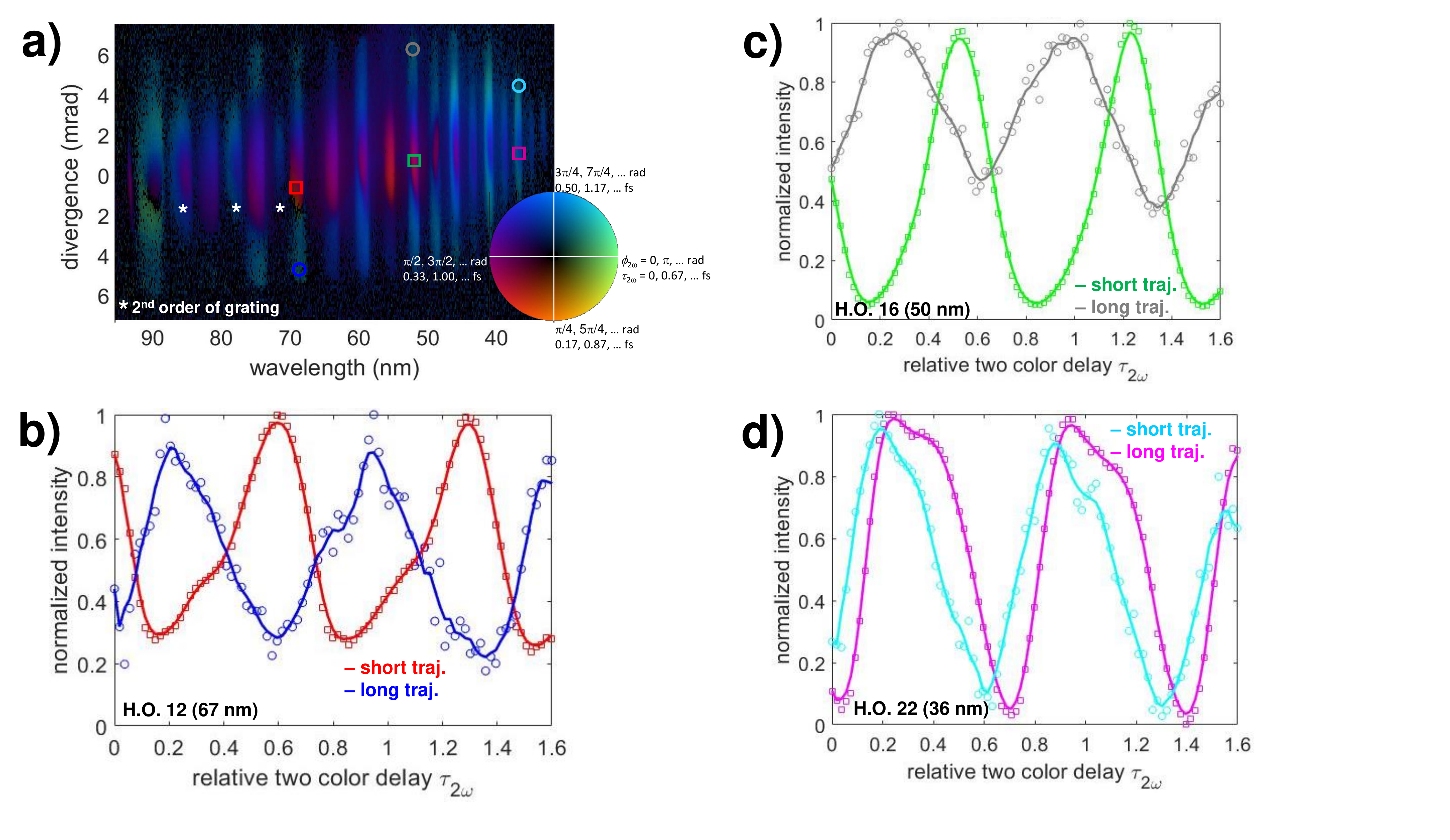}
\caption{\textbf{HHG as function of two-color phase}. (a) Amplitude (brightness) and phase (color) of a two-color phase scan of HHG. The intensity oscillations as function of two-color phase for selected regions of interest marked in (a) are shown for (b) harmonic order 12 (67 nm), (c) harmonic order 16 (50 nm), and (d) harmonic order 22 (36 nm) as function of the two-color delay $\tau_{2\omega}$ (in fs, a two-color phase shift of $\pi$ corresponds to a delay of 0.67 as, as illustrated in the legend of panel (a)). 
The open squares correspond to short electron trajectories, and the open circles correspond to long trajectories.
The open symbols in (b)-(d) are raw data, and the lines are 5-point moving averages. }
\label{fig:sol}
\end{figure*}
\\
We now turn to the results of a complete scan of the two-color phase. These results are summarized in Fig.~\ref{fig:sol} and reveal the spectral bandwidth over which two-color HHG can effectively control the divergence. For the data in Fig.~\ref{fig:sol}(a), we scanned over slightly more than two intensity oscillations. We plotted amplitude (brightness) and phase (color) of the frequency component, with which the intensity oscillates (as obtained from a Fourier Transform). Therefore, bright (as opposed to dark) regions in Fig.~\ref{fig:sol}(a) indicate a strong harmonic signal, and the color contrast indicates different oscillation phases. As explained in section 2, the signal of the short trajectories will be enhanced at a different delay than the signal of the long trajectories. The centers of most plateau harmonics (between 50 and 75 nm) show a red/purple shading (relative two-color phase of $\approx$1.2~rad, short trajectories), whereas the regions at larger divergence angles have a light blue color (phase of $\approx$2.7~rad, long trajectories). This color contrast between red and blue tones thus directly visualizes the parts of the far-field beam that are dominated by long and short trajectories, respectively.\\
Figures~\ref{fig:sol}(b)-(d) show the intensity oscillations as a function of the two-color phase in the far-field corresponding to selected harmonic orders 12 (67 nm), 16 (50 nm) and 22 (36 nm) for both short (squares) and long (circles) trajectories.
The phase between two peaks corresponds to $\pi$ (0.67 fs), such that the relative delay for maximizing the short or the long trajectory is different by approximately 0.5$\pi$ (0.33~fs). Comparing harmonic order 12 in Fig.~\ref{fig:sol}(b) and 16 in Fig.~\ref{fig:sol}(c), we note that this phase difference between the two trajectories becomes smaller for higher harmonics. This effect is even more visible for a harmonic in the cut-off region, see Fig.~\ref{fig:sol}(d). The long and short trajectories that correspond to higher photon energies are becoming more similar to each other in terms of ionization and recombination times as well as excursion amplitudes, and merge in the cutoff, which explains the similar oscillation phases for higher harmonics. 
Figure~\ref{fig:sol} thus demonstrates that we can discriminate between the long and short trajectories by adjusting the two-color phase. \\
\begin{figure*}[htbp]
\centering
\includegraphics[width=\linewidth]{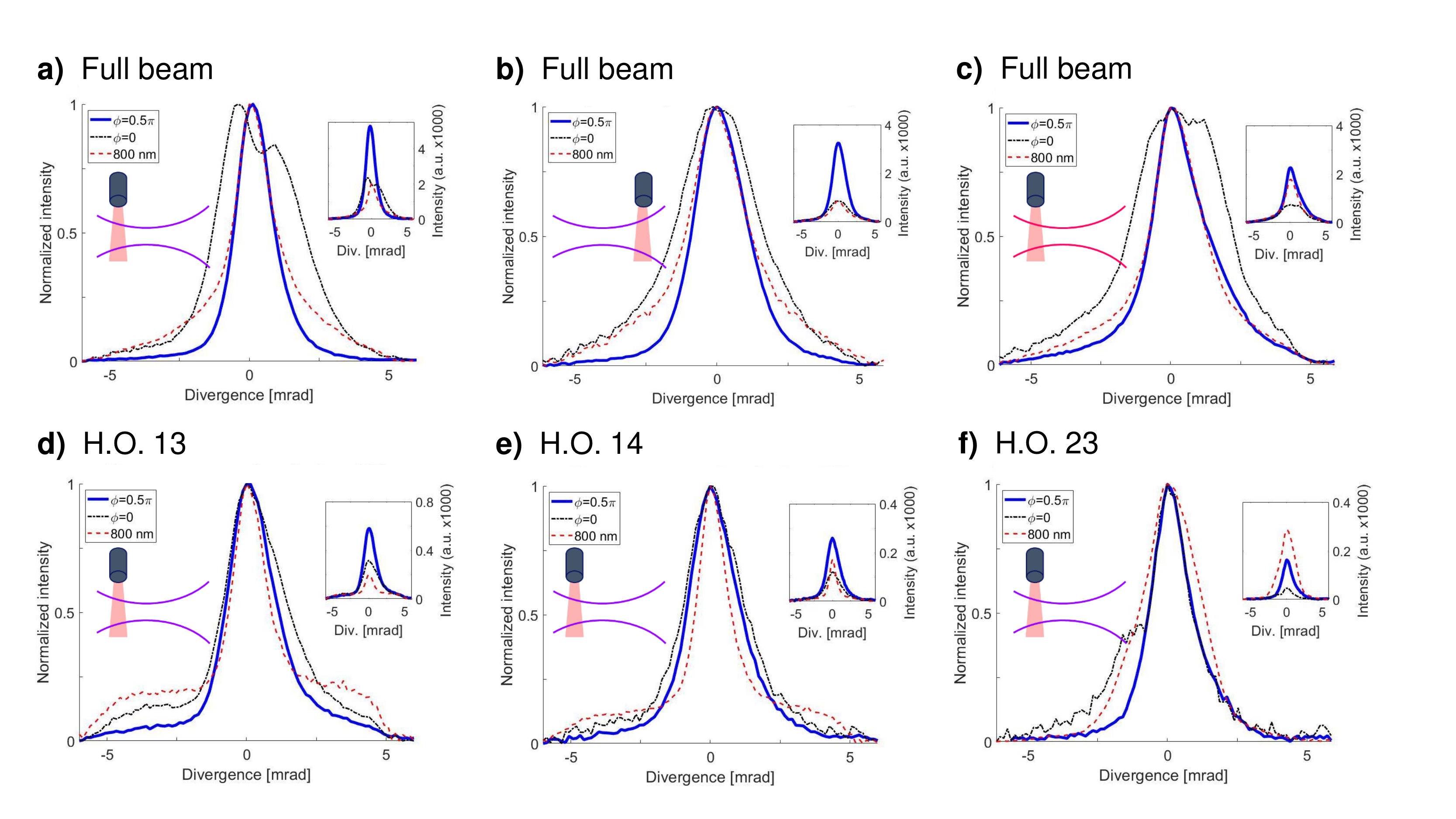}
\caption{ \textbf{Divergence control and signal enhancement in two-color HHG:} All plots in the main panel show normalized beam profiles for a two-color phase with minimized divergence (solid blue lines, corresponds to predominantly short-trajectory emission), maximized divergence (dash-dotted black lines, corresponds to predominantly long-trajectory emission) and HHG with a one-color 800 nm pulse (dashed red line). The insets show the same profiles with their actual intensities. Panel (a) corresponds to the nozzle being placed before the focus, and with 25\% relative intensity of the second harmonic. Panel (b) corresponds to the nozzle placed after the focus of the two-color pulses under otherwise identical conditions. Panel (c) corresponds to the same geometry as (a) with a weaker second harmonic at 11\% intensity. Panels (d)-(f) correspond to the same measurement as panel (a) and show individual harmonic orders 13 (62 nm), 14 (57 nm), and 23 (35 nm).}
\label{fig:div}
\end{figure*}

We now turn to analyzing and quantifying the divergence change and the signal level of the harmonics for different relative two-color delays in Fig.~\ref{fig:div}. 
The divergence of the harmonics is determined by summing the intensity along the wavelength over a region of interest. Figure~\ref{fig:div}(a) shows beam profiles, for a region of interest spanning harmonic 13 through harmonic 23. The profiles in the main panel are normalized to their maxima to highlight the effect of the two-color fields on the divergence. The intensity of the second harmonic was 25\% of the fundamental. The spectrum was generated with the gas jet before the focus, which is the configuration typically chosen to suppress long trajectory contributions \cite{salieres01a}. Nonetheless, the profile of the beam generated with the 800~nm fundamental only (red curve) is still clearly not mono-Gaussian, indicating the presence of both long and short trajectories. The blue curve displays the HHG beam profile for a relative two-color phase of 0.5$\pi$~rad, at which the short trajectories are enhanced. The black curve represents the beam profile for a relative phase of 0~rad, optimizing the contribution of the long trajectories. The beam generated with an 800+400~nm driver (blue curve) has much less signal in the wings at larger divergence angles in Fig.~\ref{fig:div}(a), compared to HHG with the fundamental 800~nm field only. Simultaneously, the total signal at this relative phase is enhanced by a factor of 2.2 compared to the HHG with the fundamental field, as can be seen in the inset in Fig.~\ref{fig:div}(a), which shows the actual and not the normalized intensity. Each beam profile can be fitted by a double Gaussian, as previously illustrated in Fig.~\ref{fig:prob}(b), which allows estimating the relative contributions from long and short trajectories. In the case of two-color HHG optimized for short trajectories (blue), the amplitude of the long trajectory contribution is decreased by a factor of 1.9 compared to the conventional one-color HHG (red). This demonstrates that for a relative phase of 0.5$\pi$~rad, both the divergence and the total signal level are improved. At this phase the instantaneous ionization rate is modified in favor of the short trajectories and the long trajectories are prevented from recombining. At a relative delay of 0~rad, displayed by the black curve, the signal in the wings is enhanced. In addition, the total signal level is comparable to the signal for the HHG with the fundamental 800~nm field only. \\
The change of the instantaneous ionization rate as function of the two-color phase is in general small for the long trajectories, and in particular less significant than for the short trajectories.
Therefore, a relative delay of 0~rad does not enhance the signal level of the long trajectories a lot. Simultaneously this phase deflects electrons related to short trajectories, thus decreasing the overall signal from the short trajectories. \\

Figures~\ref{fig:div}(d)-(f) show the beam profile changes for three separate harmonic orders 13 (62 nm), 14 (57 nm), and 23 (35 nm), taken from the same scan as the total profile shown in Fig.~\ref{fig:div}(a). In order to compare the even harmonics with the fundamental field, an average of the neighbouring odd harmonics is taken for generating the red curve in Fig.~\ref{fig:div}(e). For lower harmonics, long and short trajectories have vastly different excursion amplitudes. Therefore the low harmonics show a large difference in total intensity and divergence for the two different relative phases. In all cases the short-trajectory optimized 800+400~nm driven HHG emissions (blue curves) show less signal in the wings, compared to the 800 nm driven HHG (red curves). Also the overall intensity (insets) of the blue curve compared to the red curve increases with factors of 10 and 2 for harmonics 13, and 14 (Fig.~\ref{fig:div}(d,e)), but decreases for harmonic 23 by about a factor of 2 (Fig.~\ref{fig:div}(f)). We note that this harmonic was close to the cutoff, and the intensity of the two-color field was likely slightly lower than the 800~nm only as observed by a reduced cutoff. However, harmonics near the cutoff show less increase in absolute signal or even a decrease in general. The reason is that the relative excursion for long and short trajectories become similar for higher harmonics, and merge in the cutoff of the spectrum. In addition, the reshaped ionization rate in the two-color field favors a narrower range of trajectories (Fig.~\ref{fig:solu}), and neighboring half cycles will have very different peak intensities, giving rise to a reduced cutoff of the high-harmonic emission in every second half-cycle. These combined effects cause a reduced yield of cutoff harmonics in two-color HHG.\\

To further investigate the process of improving the brightness of the HHG beam, the position of the gas jet and the ratio between the power of the fundamental and second harmonic field are varied. Figure~\ref{fig:div}(b) shows the beam profiles in which the gas jet is positioned behind the focus. In this configuration phase matching enhances the long trajectory contribution \cite{salieres01a}. The signal in the wings is therefore suppressed even more in short-trajectory optimized two-color HHG (blue) compared to the fundamental field case (red), namely by a factor of 2.2 as compared to a factor of 1.9 in Fig.~\ref{fig:div}(a). The overall intensity of the signal is increased (inset in Fig.~\ref{fig:div}(b)) by a factor of 2.9, also surpassing the factor of 2.2 obtained in Fig.~\ref{fig:div}(a) where the gas jet was placed before the focus. \\
Figure~\ref{fig:div}(c) shows the beam profiles of an experiment, in which the percentage of the second harmonic power was lowered to 11\%. Although the total intensity is slightly improved at a relative delay of 0.5$\pi$ compared to 800 nm only, the improvement of the divergence is less pronounced with the lower second harmonic field power. By having a lower contribution of the second harmonics, the ionization rate will be changed less compared to the fundamental field only case. Also the selection of trajectories by introducing a smaller lateral velocity component will become less efficient.

\section{Conclusion and Outlook}
We demonstrated that phase-controlled orthogonally polarized two-color fields can be used to minimize the divergence of HHG, while simultaneously improving the overall flux. The improvement of the divergence is predominantly attributed to the introduction of a lateral momentum component that enables trajectory selection, while the enhanced number of photons is mainly influenced by the reshaped strong-field ionization rate in a two-color laser field. The suppressed long trajectory contribution is synonymous with a smaller and cleaner mono-Gaussian focus when refocusing the harmonics.
Such improvements will be hugely beneficial for attosecond science and lenless imaging with HHG sources.
Pump-probe experiments with two attosecond pulses from HHG, sometimes dubbed the ''holy grail of attosecond science", have made tremendous progress \cite{tzallas11a,okino15a}, but remain challenging because the peak intensities attainable with individual pulses from HHG are relatively low. All experiments that aim at attosecond pump-probe or non-linear processes in the extreme ultraviolet and soft x-ray range will benefit tremendously from our two-color strategy to achieve brighter HHG with higher-quality foci. \\
Many imaging applications need a small focus, as a too large field-of-view can go at the expense of resolution. In addition, a non-clean (mono-Gaussian) focus can cause problems in image-reconstruction algorithms that do not directly retrieve the beam, and it will always cause problems when a fine detail embedded in a larger structure is imaged and a non-mono-Gaussian focus would create imaging artifacts due to diffraction from the surrounding structures.
Improvements of the focusing of HHG are not limited to two-color fields, but to any manipulation in the HHG process that favors trajectory selection. This seems to be the beginning of a larger effort, as recent papers have highlighted the importance of understanding and improving \cite{wikmark19a,quintard19a} the micro-focusing of HHG \cite{drescher18a}.

\section*{Funding Information}
Netherlands Organisation for Scientific Research (NWO) Veni grant 016.Veni.192.254.

\section*{Acknowledgments}

This work is carried out at the Advanced Research Center for Nanolithography (ARCNL), a public-private partnership of the University of Amsterdam (UvA), the Vrije Universiteit Amsterdam (VU), the Netherlands Organisation for Scientific Research (NWO), and the semiconductor equipment manufacturer ASML. We thank Reinout Jaarsma for technical support. We thank the mechanical workshop and the design, electronic, and software departments of ARCNL for the construction of the setup. P.M.K. acknowledges support from NWO Veni grant 016.Veni.192.254.

\section*{Disclosures}

The authors declare no conflicts of interest.

 



%

\end{document}